\documentstyle[11pt,newpasp,twoside,epsf]{article}
\def\ltsima{$\; \buildrel < \over \sim \;$}
\def\simlt{\lower.5ex\hbox{\ltsima}}
\def\gtsima{$\; \buildrel > \over \sim \;$}
\def\simgt{\lower.5ex\hbox{\gtsima}}

\def\edcomment#1{\iffalse\marginpar{\raggedright\sl#1\/}\else\relax\fi}
\reversemarginpar

\begin{document}
\title{The X-ray view of highly obscured AGN}
 \author{Giorgio Matt}
\affil{Dipartimento di Fisica, Universit\`a degli Studi Roma Tre,
via della Vasca Navale 84, I--00146 Roma, Italy}

\begin{abstract}
The properties of Compton--thick AGN are reviewed, with particular
emphasis on BeppoSAX hard X--ray observations. I also discuss evidence
for the presence of Compton--thick, circumnuclear matter in unobscured
AGN, and I briefly introduce the {\sc hellas2xmm} 5--10 keV survey,
discussed in more details in other contributions to this volume.
\end{abstract}

\section{Introduction}

Both direct (surveys) and indirect (synthesis models of the Cosmic 
X--ray Background, XRB) methods 
clearly indicate that most AGN are `obscured' in X--rays, i.e. their
emission is seen through absorbing material in excess of the Galactic one.
Often (see next section) this absorbing matter is
very thick, its column density exceeding the value, 1.5$\times$10$^{-24}$
cm$^{-2}$. for which the Compton scattering optical depth equals unity.
For many years it has been customary to identify the absorber with the 
torus envisaged in popular and, on the whole, very successful Unification
Models (Antonucci 1993). We now know that the strictest version of the 
Unification Models needs modifications, and there are 
alternative unification models like
the outflowing wind proposed by Elvis (2000, and this conference). In the
following, I will call `torus' the (sub)pc--scale, large covering factor,
Compton-thick circumnuclear matter for which there is plenty of evidence in
X--rays (and in other bands), {\sl whatever its actual geometrical 
configuration}, to be distinguished from dust lanes or other distant
(scale of hundreds of parsecs), Compton--thin matter.

I have used the word `obscured', instead of `type 2', deliberately.
In the following, I will use `type 1' and `type 2' 
in their original meaning, which is 
based on the optical emission line spectrum.
As far as X--ray are concerned, I will classify sources as `obscured'
and `unobscured', the former including all sources in which there is 
absorption significantly in excess of the Galactic one. 
In the classical Unification Models, 
a one--to--one relation between optical type 1 and unobscured
sources, and between type 2 and obscured sources, is predicted. 
We now  know that there is plenty of exceptions (see below and other
contributions in this volume): type 1 sources may be obscured in X--rays, 
and X--ray selected sources may simply not appear as AGN in the optical.
Of course, I am not
claming here that there is no relation whatsoever between optical and X--ray 
appearances. More often than not, the optical (X--ray) appearance is just
what one would predict from Unification Models  after observing 
the X--ray (optical) emission.
I am only saying that sometimes obscured sources turn out to be type 1 (e.g.
Fiore et al. 2001; see also Maiolino, this volume), 
at one extreme, or dull galaxies (e.g. Mushotzky et al. 2000;
Fiore et al. 2000; Barger et al. 2001; Hornschemeier et al. 2001), 
at the other extreme. On the contrary,
I am not aware of any certainly unobscured AGN which are not 
type 1\footnote{Hardness ratios alone may not be sufficient to indicate
obscuration in case of Compton--thick absorbers; with an appropriate 
choice of the
bands, NGC~1068 would appear a very soft source indeed!}, and of
any type 2 which are not obscured. 
So, at present the `strict' (in the sense of no exceptions found yet)
relations between optical and X--ray classifications are reduced to:

\begin{center}
{\bf type 1 $\leftarrow$ unobscured}
\end{center}

\begin{center}
{\bf type 2 $\rightarrow$ obscured}
\end{center}

I would not be too much surprised, however,
if in the future exceptions to these rules would also be found.

The mismatch between optical and X--ray properties (mismatch from the 
Unification Models point of view) may solve the long standing 
problem of the lack, or at least rarity, of type 2 QSO (in fact, at least
one convincing case has been presented by Norman et al. 2001), which sometimes
is considered a problem for the standard model of the XRB. 
From X--ray surveys, it is clear that obscured, high luminosity AGN exist, 
and that is enough for explaining the XRB. 
The fact that type 2 high luminosity AGN
are not found copiousuly, if not due to some selection effects, may indicate
that, for some unclear but certainly worth studying reason, at high 
luminosity (amd/or high redshift)
the Narrow Line Regions are invisible or absent
altogether. This in turn suggests a luminosity/redshift dependence
of the structure of the absorber (as in the obscured Quasar growth model 
proposed by Fabian 1999). For instance, the absorber 
may fully cover the nucleus, so either obscuring the NLR or
preventing the ionizing photons from illuminating them.

\section{General properties of X--ray absorbers}

X--ray absorption is very common in AGN.
All Seyfert 2s observed in X--ray are absorbed, and because Seyfert 2s
outnumbers Seyfert 1s by a factor of a few, at least in the local Universe,
this implies that {\sl optically selected} AGN are preferentially obscured.
Synthesis models of the XRB (e.g. Setti \& Woltjer 1989; 
Comastri et al 1995; Comastri et al. 2001;
see also Comastri 2001 for a review, and the references therein for the
several flavours of the model) also require a large fraction 
of absorbed sources. Heavy absorption is also
very common: about half of the optically selected Seyfert 2s in the local
Universe are Compton--thick (e.g. Maiolino et al. 1998). Indeed, the very
first object observed by XMM--$Newton$ in the framework of a program devoted
to study the absorption properties of optically selected Seyfert 2s, NGC~4968,
is clearly a Compton--thick source (Guainazzi et al 2001),  as shown
in Fig.~1. Actually, there is a relation between optical 
classification and column density: Risaliti et al. (1999) have in fact
shown that Intermediate (1.8-1.9) Seyferts are usually Compton--thin,
while classical Seyfert 2s are Compton--thick. It is possible (Matt 2000)
that  Compton--thin/Intermediate Seyferts are obscured by the dust lanes
at distances of hundred of parsecs which Malkan et al. (1999) found to be
common in Seyfert galaxies\footnote{Compton--thin absorption may also be partly
due to matter much closer to the Black Hole like the BLR, see Risaliti et al. 
2001 for evidence based on
variability studies.}, while Compton--thick/Seyfert 2s are obscured
by the torus (in the abovementioned meaning). The heavy absorbers
should have a large covering factor, to account for the large
fraction of Compton--thick sources, and must be fairly compact, not to 
exceed the dynamical mass, at least in the best studied cases like
Circinus and NGC~1068 (Risaliti et al. 1999). (For these two sources, a completely
independent estimate of the inner size of the torus has been derived by
Bianchi et al, 2001 and this volume, by modeling the X--ray line spectra.
They found a minimum distance of the torus of about 0.2 and 4 pc,
respectively). 

\begin{figure}
\plotfiddle{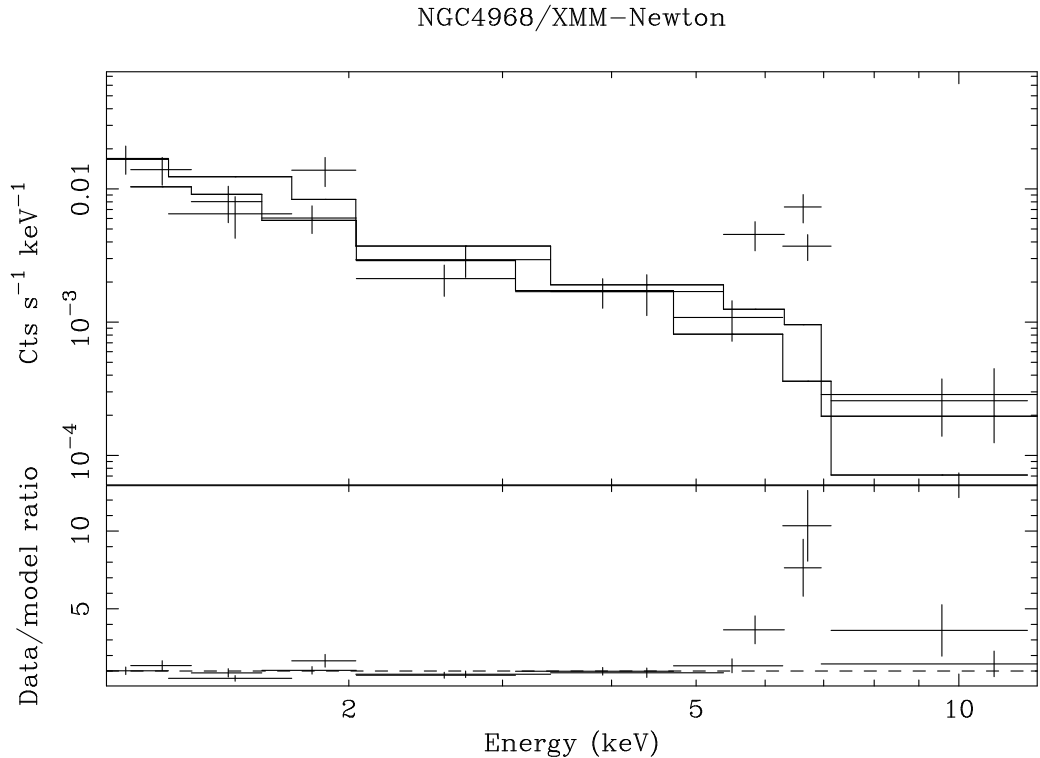}{10cm}{-90}{58}{58}{-220.}{330.}
\caption{ The XMM--$Newton$ spectrum of the Seyfert 2
galaxy NGC~4968 (Guainazzi et al. 2001). The spectrum is well fitted  by a pure cold reflection
component plus a prominent iron line, indicating a Compton--thick source.}
\end{figure}

\section{The X--ray spectrum of bright Compton--thick AGN}

Just because they are so heavily absorbed, Compton--thick AGN are the
ideal sources for studying  the circumnuclear matter. The
reflection components which, in unabscured sources, would be diluted into
invisibility by the primary radiation, are here well visible.  In these
sources absorption is so heavy that no nuclear radiation is observable
in the `classical' X--ray band, i.e. below 10 keV. If the column density
is of the order of a few times
10$^{24}$ cm$^{-2}$, however, the nucleus may become visible at energies
of a few tens of keV (see Fig~2). At these energies, the most sensitive
instrument so far has been the PDS onboard BeppoSAX.

\begin{figure}
\plotfiddle{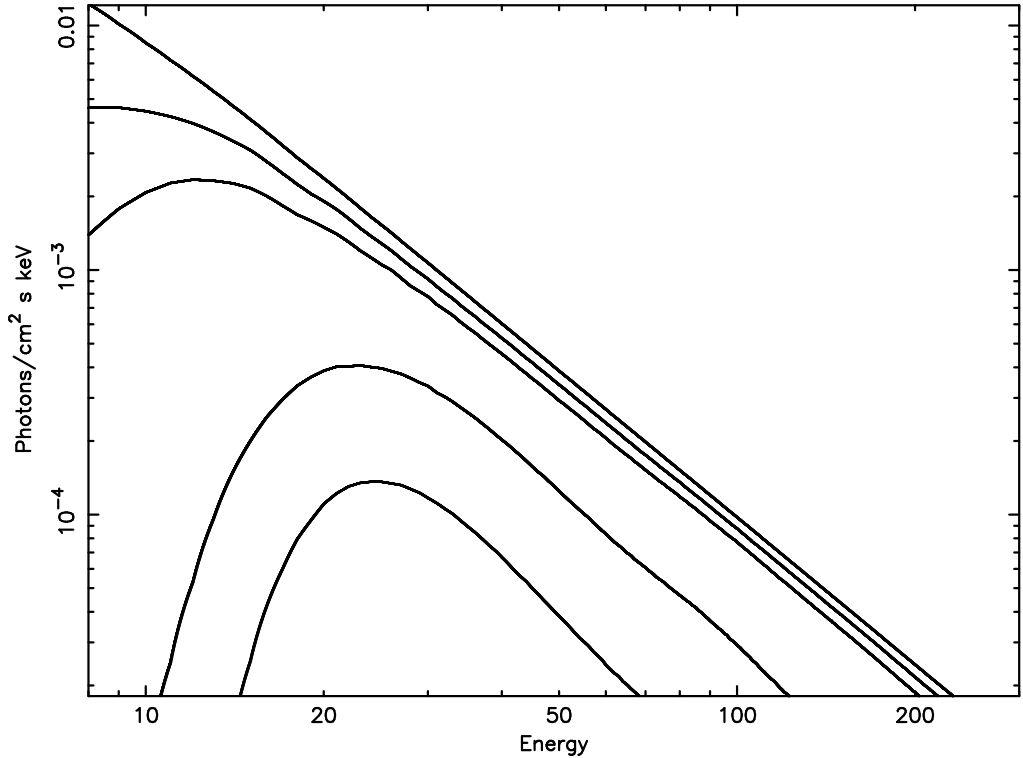}{10cm}{-90}{58}{58}{-220.}{330.}
\caption{The effect of absorption on the X--ray spectrum of AGN
(assumed to be a power law with photon spectral index of 2). Curves are for
$N_H$=10$^{23}$, 5$\times10^{23}$, 10$^{24}$,  5$\times10^{24}$ and
 10$^{25}$ cm$^{-2}$ (from top to bottom). 
Spectra have been computed following  Matt et al. (1999), and include 
the effects of Compton scattering in a spherical geometry.}
\end{figure}

\begin{table}
\begin{center}
\begin{tabular}{||l|c|c|c|c||}
\hline
& & & & \cr
Source &   $N_H^a$ &   CR &   WR & 
  $L_X^b$ \cr
 & & & & \cr
{NGC~1068$^{1,2}$} &   $\simgt$10 &   Y &   Y &   ?~($>1$) \cr
{Circinus Galaxy$^{2,3}$} &   4.3 &   Y &   N~(?) &   $\sim$0.01 \cr
{NGC~6240$^4$} &   2.2 &   Y~(?) &   Y &   $\sim$1.2 \cr
{Mrk~3$^{5}$} &  1.1 &  Y & Y?  & 0.9  \cr
{NGC~7674$^{6}$} &  $\simgt$10   & Y & N~(?)  & ?  \cr
{NGC~4945$^7$} &  2.2 &  N & N & $\sim$0.03  \cr
{TOL~0109-383$^8$} &  2.0 &  Y &   Y &   $\sim$0.2 \cr
& & & & \cr
\hline
\end{tabular}
\end{center}
\caption{Summary of the main properties of bright Compton--thick AGN.
CR and WR stand for Cold Reflection and Warm (i.e. ionized) 
Reflection, respectively. Notes: a)in units of 10$^{24}$~cm$^{-2}$; 
b) 2--10 keV luminosity 
in units of 10$^{44}$~erg s$^{-1}$. References:
1) Matt et al. 1997; 2) Guainazzi et al. 1999; 3) Matt et al. 1999; 
4) Vignati et al. 
1999; 5) Cappi et al. 1999;  6) Malaguti et al. 1998; 7) Guainazzi et al.
2000; 8) Iwasawa et al. 2001a}
\end{table}

In the following subsections I will
describe in some details the BeppoSAX results for the brightest of them
(for a summary, see Table 1).
For other, fainter Compton--thick AGN observed by BeppoSAX see:
Malaguti et al. (1998) for NGC 7674;  Iwasawa et al. (2001a) for 
Tololo 0109-383; Franceschini et al. (2000) for IRAS 09104+4109. The
data of Mrk~3 are presented by Cappi et al. (1999), while the interesting
case of Arp~220 is discussed by Iwasawa et al. (2001b, and this volume).
For a general discussion of BeppoSAX results on  Compton--thick
sources see Matt et al. (2000). 

\subsection{NGC~1068}

NGC~1068, the archetypal Seyfert 2 and Compton--thick galaxy, has a very
complex X--ray morphology, as revealed by $Chandra$ observations 
(Young et al. 2001). The bulk of the iron K$\alpha$ lines, however, come from
the nucleus. ASCA (Ueno et al. 1994; Iwasawa et al. 1997; Netzer \& Turner
1997) resolved three iron K$\alpha$ lines, corresponding to cold, He-- and
H--like iron. Matt et al. (1996) and Iwasawa et al. (1997) suggested that 
at least two reflectors were at work, one cold and optically thick (to
be identified with the torus), the other hot and optically thin. Matt
et al. (1997) confirmed these suggestions by disentangling the two reflecting
continua in the BeppoSAX spectrum. No direct emission is observable, 
suggesting a very thick ($N_H$ of 10$^{25}$ cm$^{-2}$ or more) absorber.
A complete analysis of the BeppoSAX
data revealed also a O {\sc vii} line (Guainazzi et al. 1999, now confirmed
by $Chandra$, Young et al. 2001, and XMM--$Newton$, Kinkhabwala et al. 2001) 
which, combined with the lines from
intermediate $Z$ elements like Mg, Si and S, implies that at least a
third, warm reflector is present. Bianchi et al. (2001 and this volume) 
reanalysed the BeppoSAX and ASCA data, and estimated the physical and 
geometrical properties of the reflectors. In particular, the inner surface of 
the torus was estimated to be at a distance of $\sim$4 pc from the Black Hole.

Comparing the two BeppoSAX observations, performed about 1 year apart,
Guainazzi et al. (2000) measured an energy dependent flux variation, 
explained in terms of a variation of one or both
the ionized reflectors. This would imply a size, for these reflectors,
of the order of magnitude of a parsec or so.

\subsection{Circinus Galaxy}

Also for the Circinus Galaxy, $Chandra$ observations (Sambruna et al.
2001a,b) show a rather complex morphology, but the bulk of the reflection
comes from the nuclear region. The spectrum below 10 keV is dominated
by a cold reflection component (Matt et al. 1996), while the nuclear
radiation becomes visible at energies of a few tens of keV (Matt et al.
1999). An analysis of the ASCA and BeppoSAX line spectrum (Bianchi et al.
2001 and this volume) suggests that the inner surface of the torus 
is at a distance of about 0.2 pc from the Black Hole. 

\begin{figure}
\plotfiddle{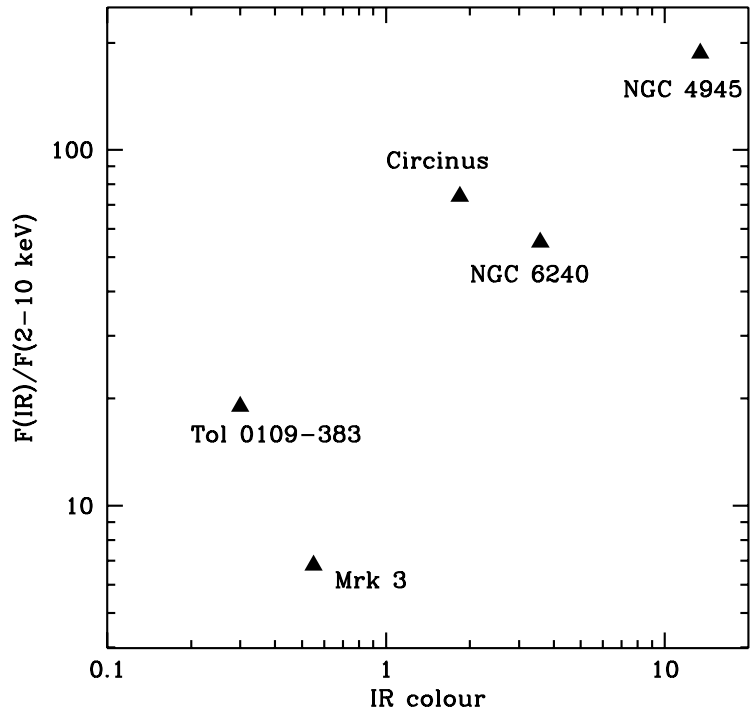}{10cm}{-0}{60}{60}{-180.}{-100.}
\caption{The ratio between the IR  and 2--10 keV fluxes (corrected
for absorption) vs
the IR colour (defined as:
$(S_{60}\nu_{60} + S_{100}\nu_{100})/(S_{12}\nu_{12} + S_{25}\nu_{25})$),
for the sources in the sample of Matt et al. (2000) for which 
an estimate of the nuclear X--ray flux is possible.}
\end{figure}

\subsection{NGC~6240}

Since the discovery by BeppoSAX of the nuclear radiation piercing through
a Compton--thick absorber (Vignati et al. 1999), NGC~6240 has became the
archetypal obscured, high--luminosity source, and its SED is widely used
to study the relations between IR and X-ray surveys and cosmic bacgkrounds.
From IR diagnostic tools,
the source appears to be dominated by starburst emission (Genzel et al.
1998), but already Iwasawa \& Comastri (1998) found strong evidence for
AGN activity in the ASCA spectrum. This was definitely confirmed by the
BeppoSAX observation cited above. The X--ray nuclear luminosity is rather
large and, when translated to the bolometric luminosity, suggests 
that the AGN dominates the energy output in this source.
The pretty cold IRAS colours, however (see Fig.~3),
point to starburst emission for the far IR. It is likely that in this 
source starlight and accretion powers are
actually of the same order of magnitude. Whether this is just
a coincidence, or it is indicating a strong relation between the two
phenomena, it is not clear at the moment, but it is certainly a very
intriguing question.

Finally, it is worth remembering that from optical spectroscopy NGC~6240 is
classified as a LINER (Veilleux et al. 1995), representing one of the
first example of the mismatch between optical and X--ray classifications,
for which there are now several cases from $Chandra$ and XMM--$Newton$ 
surveys.

\subsection{NGC~4945}

An even more extreme case of mismatch between optical and X--ray 
classifications is that of NGC~4945, for which there is no evidence
whatsover for AGN activity at all wavelenghts but in hard X--rays. 
Iwasawa et al. (1993) discovered a heavily obscured nuclear emission
in the $GINGA$ spectrum of this source. Later on, Done et al. (1996) confirmed
this result with a RXTE observation. Recently, Guainazzi et al. (2000)
and Madejski et al. (2000), from BeppoSAX and RXTE observations respectively,
discovered large amplitude variations on time scale of thousands of seconds
in hard X--rays, where the nuclear radiation dominates the emission. 
With the column density of the absorber, most of the radiation should 
emerge after
one or more scatterings, if the covering factor of the absorbing matter is
large, and no short term variation should be visible, unless the absorber
is very close to the nucleus (of the order of milliparsecs or less) or,
more likely, the covering factor of the absorber is in this source,
differently from the common case, rather small.

\section{Evidence for the `torus' in unobscured AGN}

\begin{figure}[t]
\plotfiddle{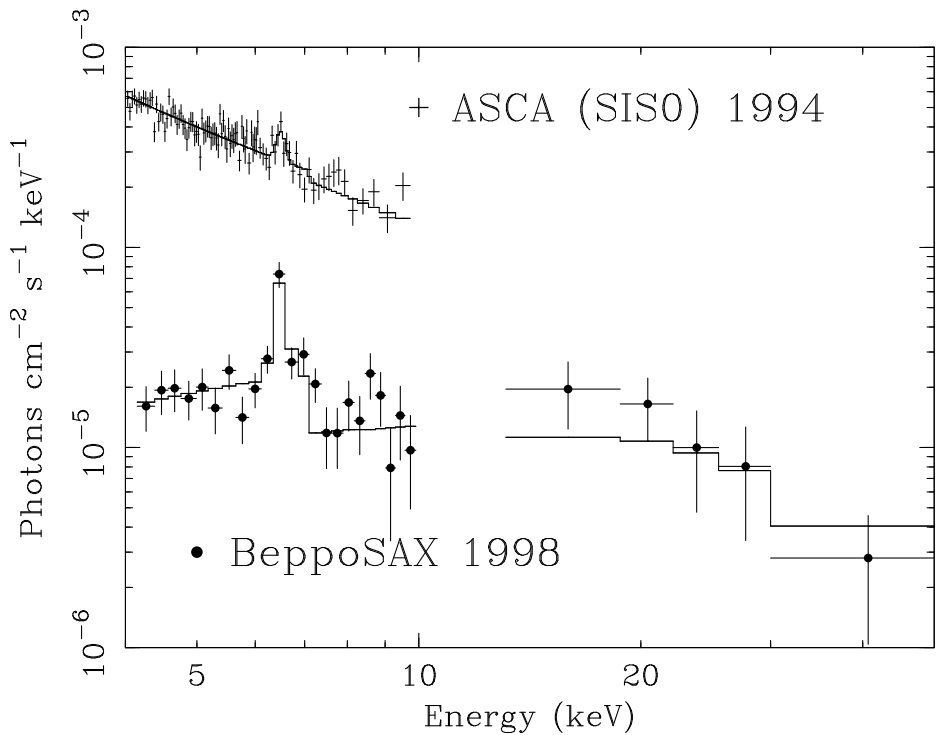}{10.cm}{-90}{56}{56}{-220.}{330.}
\caption{The 1994 ASCA and 1998 BeppoSAX spectra of NGC~4051. During the 
BeppoSAX observation, only the reflection component was visible.
Courtesy of Matteo Guainazzi.}
\end{figure}

One of the least tested prediction of Unification Models is that the torus
should be present also in type 1/unobscured objects. One of the best way 
to check it is to search for X--ray reprocessed components
(Ghisellini et al. 1994, Krolik et al. 1994). The problem is that reprocessing 
also occurs in the accretion disc, but fortunately there are at least two
significant differences between the two cases: 
firstly, reprocessed components from
the accretion disc should respond to variations of the primary emission
on very short time scales (thousands of seconds), while those from the
torus should be delayed by months or years, due to the far larger
distance from the Black Hole. Secondly, features originating in the innermost
part of the accretion disc are strongly affected by GR and kinematic effects,
contrary to those from the torus. In this respect, iron line profile is
much more useful than the Compton reflection continuum, modifications
from the GR effects being much more dramatic (Fabian et al. 1989; Matt
et al. 1991; Fabian et al. 2000 and references therein). 

The most convincing case so far for a type 1 AGN with the torus has been
obtained by variability studies. In 1998 NGC~4051 went to a prolonged
(several months) low state (Uttley et al. 1999), 
almost at the end of which was observed by BeppoSAX (Guainazzi et al. 1998). 
From the spectrum, it was clear that the primary emission was disappeared
into invisibility, while the reflection components were still there. 
(This is best seen in comparison with a previous ASCA observation 
performed during a normal activity state, see Fig.~4). Reprocessing
should therefore occurs in matter at a distance of at least a
few light--months from the Black Hole. It is worth noting that the
reprocessing material must be Compton--thick, to provide a significant
reflection continuum.

Observations of the iron line profiles are more ambiguous. $Chandra$ gratings
and XMM CCDs are discovering several narrow iron lines in Seyfert 1s 
(e.g. Kaspi et. al. 2001; Yaqoob et al. 2001; Reeves et al. 2001; Gondoin
et al. 2001; Pounds et al. 2001; see also Lubinski \& Zdziarski 2001 for 
a reanalysis of ASCA data). Without a determination of the amount
of associated reflection continuum (hard to achieve with $Chandra$ and 
even with XMM--$Newton$, due to the limited bandwidth)
it may be difficult, however, to
understand if these narrow lines come from the torus or from Compton--thin
matter, like the Broad Line Region. Even the $Chandra$
gratings may not always be sufficient to resolve the line at the 
velocity of the BLR, often for want of photons rather than of spectral
resolution. To overcome these problems, simultaneous observations with
hard X--ray instruments have to be performed. We observed NGC~5506
(a Compton--thin obscured AGN, and therefore a source for which
there is no direct evidence for the torus) 
simultaneously with XMM--$Newton$ and BeppoSAX (Matt et al. 2001). 
The iron line, as measured by the EPIC/p-n instrument onboard
XMM--$Newton$, is clearly complex, being composed by at least two
components: one unresolved and corresponding to neutral matter, 
the other broad and corresponding to ionized matter (see Fig.~5).
Combining the p-n spectrum with the BeppoSAX/PDS one, a strong reflection
component is clearly detected. This component clearly comes from
neutral matter (as derived from the edge--like feature at 7.1 keV
observed in the p-n spectrum) and must therefore be associated with
the narrow line. Both components, henceforth, are originated in distant,
Compton--thick material.

\begin{figure}[t]
\plotfiddle{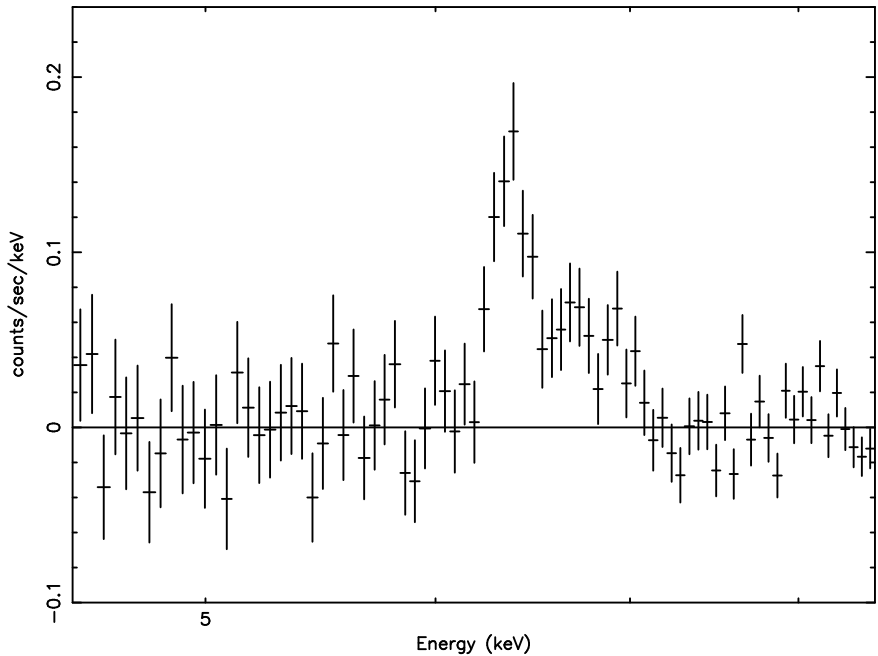}{9cm}{-90}{50}{50}{-220.}{300.}
\caption{ The iron line complex in NGC~5506 from an XMM--$Newton$
observation (Matt et al. 2001).}
\end{figure}

\section{Obscured AGN in the Universe}

When looking at the Spectral Energy Distribution of the extragalactic
Backgrounds, the XRB may appear almost negligible when compared with 
the sub--mm and IR backgrounds. One can therefore wonder why bothering to
perform X--ray surveyes, as they seem to regard a tiny fraction of the
energy in the Universe. There are at least two answers to this question.
Firstly, the X--ray band is by far the cleanest band where to
study accretion luminosity: any point--like
source with an X--ray luminosity exceeding 10$^{42}$ erg s$^{-1}$ or so
is very likely powered by accretion. 
Secondly, the luminosity we observe in the XRB is only a minority,
probably of the order of 10\%, of the energy really emitted, the
remaining 90\% having been absorbed and reemitted at longer wavelenghts.
For instance, Fabian \& Iwasawa (1999) estimated that a by no means
negligible fraction (something like 20\% or so) of the IR background is
actually due to X--ray photons absorbed and reprocessed from the obscuring
medium. 

Just because most AGN are absorbed (and, moreover, most of the absorbed
AGN are Compton--thick), the ideal survey would be at 30 keV (where the
XRB peaks!). Unfortunately, at these energies
no real imaging instrument is available until
the launch of Constellation--X at the end of the decade. In the meanwhile, 
we are forced to use the hardest imaging instrument available at present.

Prior to the launch of XMM--$Newton$, the MECS instrument onboard BeppoSAX
provided the best compromise between hard X--ray sensitivity and spatial
resolution. We therefore started a project called {\sc hellas} (High Energy
Large Area Survey; Fiore et al. 1999; 2001) aiming to search and identify 
sources in the 5--10 keV energy range. It is important to note that such
a hard selection not only exclude most non AGN X--ray sources, but
discovers sources not observable (because heavaly obscured) in soft
X--ray surveys (a number of {\sc hellas} sources have in fact not been 
detected by ROSAT, see Vignali et al. 2001). The {\sc hellas} survey has
resolved about 25\% of the 5--10 keV XRB (Comastri et al. 2001). The
sample includes 118 sources, 74 of which have been spectroscopically
identified in the optical (La Franca et al. 2001 and this conference). The
largest fraction turned out to be type 1 AGN. This is mainly because at
these energies, and at these flux limits, unobscured objects still provide
the majority of sources. Interestingly, however, some of the objects
optically identified as type 1 are indeed obscured in X--rays 
(Fiore et al. 2001). 

The launch of $Chandra$ and XMM--$Newton$ provided of course a dramatic
improvement in X--ray surveys. Many large collaborations are active in
this field (see Hasinger, this conference)
with the final aim to understand the birth and evolution of 
Supermassive Black Holes, and their relation with galaxies. We decided
to exploit the good sensitivity of XMM--$Newton$ up to about 15 keV to
continue our project started with BeppoSAX and 
build up a sample of 5--10 keV  XMM--$Newton$ selected sources 
(Baldi et al. 2001 and this conference). We decided to prefer large area
to depth, in order to have a sizable sample of relatively bright sources,
for which the redshift and optical classification could be easily obtained
from 4m class telescopes, and for which at least basic spectral informations
can be obtained in X--rays too. First results from this project (named
{\sc hellas2xmm}) can be found elsewhere in this volume (see contributions
by Baldi and Fiore). 

\bigskip
\noindent
{\it Acknowledgements} It is a pleasure to thank all my collaborators
in these researches: S. Bianchi, A.C. Fabian, F. Fiore, M. Guainazzi, 
K. Iwasawa, G.C. Perola and all members of the {\sc hellas} team. 
I thank R. Maiolino, A. Marconi,
and M. Salvati for scientific discussions and, of
course, for having organized this conference. I acknowledge financial 
support from ASI and MURST under grant {\sc cofin-00-02-36}.

\end{document}